      [1999/12/01 v1.4c Il Nuovo Cimento]
\documentclass{cimento}
\usepackage{graphicx}    
\title{Atomic Parity Violation. Early days, present results, prospects}
\author{M.-A.~Bouchiat} 
\instlist{\inst Laboratoire Kastler Brossel, CNRS, UPMC, ENS, 24 rue Lhomond FR-75005 Paris, France}
 \PACSes{\PACSit{11.30.Er, 32.90.+a} 
  {\ldots}}
\begin{document} 
 
\maketitle

\begin{abstract}
 This is a personal recollection of the time when the search for APV was beginning. In spite of today's remarkable results, summarized here, there are still important  goals to be achieved. I indicate a possible way to tackle the remaining  experimental challenges, by adapting methods now of frequent use in precision metrology.  
\end{abstract}

\section{The early days}
 As is well known from the community, T.D.~Lee and C.N.~Yang were the first to express doubts about the conservation of Parity in weak processes. They also indicated how significant experimental tests should be performed. 
 Two systems initially symmetric are allowed to evolve under the interaction to be tested. If there exists a left-right (L-R) asymmetry in the final state, Parity is violated.  One of the experiments they suggested was performed the following year (1957)  by  Chien Shiung~Wu, who reported a nearly maximum asymmetry in the $\beta$-disintegration of polarized $^{60}$Co nuclei!

This was a great shock in the domain of concepts affecting the whole physicist community. Nevertheless, by the end of the 60's, the atomic physicists did not feel directly concerned. An atomic preference between left and right was looking extremely unlikely. An atom exhibits a high degree of  symmetry and is governed by electromagnetic (EM) interactions well-known for making no distinction between left and right. However, in the mid  1970's considerable efforts were  being engaged to test parity conservation (PC) in atoms: after all an atom is not a purely EM system, what can be  expected from the other interactions, in particular weak interactions? Could they perturb the electron orbitals when they approach the nucleus? If so, PV might occur! Things went so fast that by the end of the 70's, when experiments reported the $\underline{\mathrm{absence}}$ of PV effects in atomic experiments, there was considered to be a weak link in the chain of theoretical reasoning... But a few years later several groups were able to confirm, in a few atoms, the existence of atomic PV.  Such results complement the spectacular successes of the Standard Model (SM) unifying EM and weak theories, which  culminated with  the discoveries of the W's and Z$_0$ in high-energy experiments. Indeed, atomic experiments, performed at much lower energies, test long distances and provide information of a different nature.

The sudden motivation for testing PV in atoms resulted from the revolutionary ideas which accompanied the emergence of the electroweak theory, accounting for both EM and weak interactions in a single mathematically coherent theory. Before, weak processes were considered as being mediated only by the charged $W_{\pm}$ bosons, which make the atom unstable, so that it was taken for granted that W interactions and their associated PV feature  were not relevant to the physics of stable atoms.
Later, the crucial point became the theoretical prediction for the existence of a third heavy gauge boson, carrying no electric charge, the $Z_0$ which mediates a weak force of a novel kind, so-called Neutral Current (NC) interaction. In this case, there would no longer be anything to prevent the electron from  feeling its effect inside an atom.
However, when the search in atoms started (at ENS in 1973),  the discovery of NC at Gargamelle still looked extremely hypothetical and raised many questions: did NC  exist with neutrinos or only with charged particles? Bearing in mind the size of the interaction range, only $10^{-7 }$ times the atomic radius, would the effect, if ever present in atoms, be much too small to be detected?  

In atoms the L-R asymmetries ${\cal A}_{LR}$ are extremely tiny, no more than $10^{-6 }$, just nothing in comparison with the asymmetry measured in Mrs Wu's experiments! The reason is that the electron-nucleus $Z_0 $-exchange always competes with photon exchange, whose probability amplitude is stronger by many orders of magnitude. Thus their interference leads to such small PV asymmetries, in great contrast with $\beta$-decay which involves the PV and the PC {\it weak amplitudes} of similar magnitudes. 
Fortunately, there are large enhancement effects (pointed out in\cite{ref:bou74}): i) the growth of the asymmetry in heavy atoms, is faster than $Z^3$ and 2) it looked possible to work on highly forbidden transitions where $A_{EM}$ is strongly inhibited. The $6S_{1/2}-7S_{1/2}$ cesium transition combines both advantages. Even so, the transition rate is so small   $10^{-6}$  s$^{-1}$ that problems with the counting rate could be anticipated.  To overcome this difficulty, we  proposed the Stark interference method \cite{ref:bou74} which turned out to be successful: the applied $E$-field participates to the PV signature and its magnitude is easily adjusted to optimize the experimental conditions.

Claude and I, started to investigate the Cs suggestion, theoretically, in the fall of 1972. In june 1973, I obtained an order be passed to Spectra-Physics for their newly commercialized  color-center laser (the first in France!).  A presentation of the project was first given at a Trieste Conference by Claude in presence of Salam, who made encouraging comments. Then, with Lionel Pottier, who had just completed his thesis and his military service, we built  the first experience. Progress were done against all the odds (at first the laser delivered no light at 539 nm, the Cs  transition wavelength...
This was just the beginning of a long struggle to make it work and construct two other essential devices: the high-purity polarization modulator and the multipass Cs cell with internal mirrors.)

 We succeeded to validate the Stark interference method in 1976, by applying it to the measurement of the Cs forbidden $M_1$ amplitude \cite{ref:pot}.
 In July 1976, an animated session took place at the Atomic Physics Conference in Berkeley. I discovered that Gene Commins and his graduate student, Steve Chu, had started an experiment on the Tl $6P_{1/2}-7P_{1/2}$ $M_1 $ line, also among our initial proposals \cite{ref:bou74}. Big efforts to search for an optical rotation signal on allowed $M_1$ transitions in Bi were also reported. Several tests in hydrogen and muonic atoms were proposed with a few of them already underway.
The field was gathering momentum.  
Moreover, Jean Brossel Head of the lab, who made an inquiry to make up his opinion about  our project, received very encouraging comments from Steve Weinberg. In his answer, he  insisted on the need for experiments proving that  NC interaction does take place at the expected level between electrons and nucleons and showing whether they violate Parity, ``since this would immediately rule out the vector model". 
However, before the end of that same year, absence of any PV effect in Bi was announced by two groups... though it did not deter the other groups' efforts.
\begin{figure}
\centering\includegraphics[height=3.8 cm]{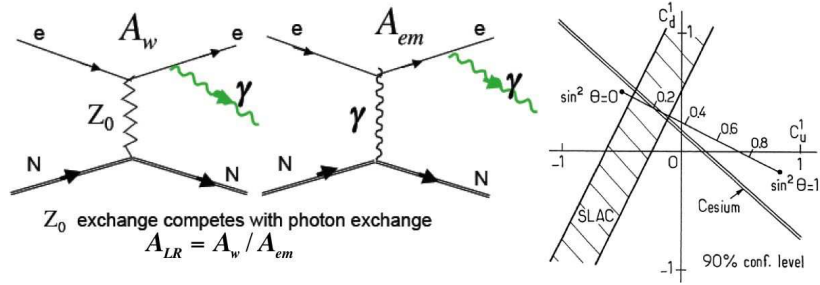}
 \caption{Left: the two amplitudes $A_w $ and $A_{em}$ which interfere in atoms and give rise to APV; Right: domains delimited in the $C_u^1$, $C_d^1$ plane by APV results in Cs and SLAC results \cite{ref:sci86}.  } \label{band}
\end{figure}

In September 1979 an important workshop was organized in Carg\`ese, by W. Williams, who passed away far too early. This was the very first time that our two communities, Atomic Physicists and High Energy Physicists involved in parity violating electron scattering (PVES),  met together \cite{ref:Will}. Charles Prescott reported the first observation of a PV asymmetry in inelastic polarized-electron scattering on deuterons at high energies, with 10$\%$ statistical accuracy and negligible systematics.  Another event was the participation of our Russian colleagues not allowed to travel before...The three Bi-groups presented positive results, some of them still preliminary. From lively discussions it emerged that the first observation of a manifestation of weak interaction in a purely atomic process had been achieved. But there was still a disturbing factor of 2 of discrepancy between the two groups in Seattle and Oxford and the Novosibirsk group.
 In Tl, a 2-$\sigma$ effect was reported,  followed by a 3-$\sigma$ one the following year \cite{ref:Com}, (see table 1 in \cite{ref:sci86}). Soon, our efforts gained driving force with the arrival of  Jocelyne Gu\'ena, as PhD student  but later a mainstay of the group, and Larry Hunter, as Post-Doc from Pr. Commins' group. 

Our first Cs result came next in 1982 \cite{ref:exp}, but it arrived with a 6$\sigma$ statistical accuracy and a very detailed analysis of systematic uncertainties, obtained by recording during data acquisition all the instrumental imperfections which may contribute to detrimental effects \cite{ref:exp}. Once this result combined with that of a second measurement obtained in 1983 on a different hyperfine component, and interpreted using atomic calculations, (then 12$\%$ accurate), 
it was becoming possible to make a comparison with the SM model prediction. This was the first quantitative test of this model at low energies. 
Its complementarity with the SLAC experiment is illustrated on fig.\ref{band}.
  
\section{Present results}
\begin{figure}
\centering\includegraphics[height=5.5cm]{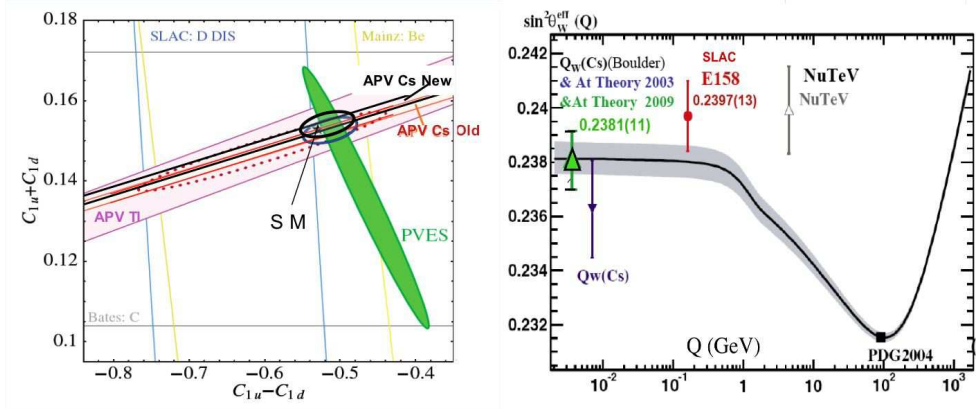}     
\caption{Left: Regions allowed by APV and PVES measurements for the weak charges of the quarks at 1$\sigma$. The black contour (95$\%$ CL) indicates the full constraint imposed by combining the results. The new Atomic Physics calculation \cite{ref:Der09} pushes the Cs band and the black contour  slightly upward. The star indicates the SM prediction. (The red dotted ellipse is the full constraint anterior to PVES analysis.) {\it Adapted from \cite{ref:You05}, with Cs band redrawn according to \cite{ref:Der09}}. Right: Running of the electroweak coupling at low energies. {\it Adapted from \cite{ref:slac}}} \label{PVES}
\end{figure}
In atomic physics,
the $Z_0$ electron-nucleon exchange is responsible for an additional term in the atomic Hamiltonian, conveniently written in the non-relativistic limit as:
$$ V_{pv}= \frac{Q_W G_F}{4\sqrt{2}} \delta^3(r_e)\frac{\sigma_e\cdot p_e}{m_e c} + H.C.   $$
In this expression the last factor is the axial-vector charge of the electron; the delta function results from the large mass of the $Z_0$, hence the short range of the interaction, causing effects  in atoms to be such small; $G_F$ is the Fermi constant and $Q_W$ plays the same role as  the nuclear electric charge in the Coulomb interaction: it is called the weak charge of the nucleus. It is the sum of the weak charges of all the constituants. This is the electroweak parameter that all APV experiments seach for. All of them rely on the measurement of the $E_1^{pv}$ amplitude of a transition not allowed by EM interactions alone.
The magnitude $E_1^{pv}$ depends not only on $Q_W $ but also on an atomic factor involving the atomic wave functions at the nucleus where $V_{pv}$ contributes, but also outside the core where the coupling with the radiation field takes place. Therefore, an atomic physics calculation is necessary to extract $Q_W$ from $E_1^{pv}$, which leads to a second source of uncertainty on $Q_W^{ex}$ (see M. Safronova, this conference).
In addition, we have to stress that this is the absolute value of $E_1^{pv}$ which has to be theoretically interpreted, while the quantity   measured by the Stark interference method is actually the asymmetry $A_{LR}= E_1^{pv}/\beta E$,  $\beta E$ being the amplitude induced by the applied electric field $E$. The uncertainties on both $\beta$ and $E$  are difficult to guarantee to better than $1\%$. For Cs, we have proposed \cite{ref:cal} to rely on one contribution to the magnetic dipole transition amplitude $M_1^{hf}$, much more precisely known ($\sim 0.1\%$) on theoretical grounds, which can be  isolated experimentally without ambiguity. Thus, 
by completing the measurement of $A_{LR}$ by one of  $M_1^{hf}/\beta E$, one can achieve absolute calibration of  $E_1^{pv}$ exceeding 1$\%$ accuracy.  

The most accurate experiment in cesium is that of the Boulder group \cite{ref:Wie}. It results from measurements of $ E_1^{pv}/\beta$ performed on the two hyperfine $\Delta F= \pm 1$ lines, combined together for eliminating the nuclear-spin dependent contribution, and from the measurement of $M_1^{hf}/\beta$ made as advocated before. Once $Q_W^{ex}$ extracted, the authors claimed 0.35 $\%$ and 0.4 $\%$ experimental and theoretical uncertainty. They used the atomic theory available at that time and they concluded to a deviation of $Q_W^{ex}$ with respect to the SM prediction $Q_W^{th}$ by 2.6 $\sigma$.  This result, published in 1999, prompted several theorists to reconsider the problem. A review of the abundant theoretical work which has followed, can be found in A. Derevianko's publication \cite{ref:Der09} which also reports on an important improvement in  the many body calculation (actually a real {\it tour de force}) leading to a final theoretical uncertainty of only 0.27$\%$, hence smaller than the experimental one,  with a readjustment of the central value of  similar size. The implications of this new result  $Q_W^{ex}= -73.16(29)_{ex}(20)_{th}$, 
are best illustrated by fig.\ref{PVES},  borrowed from \cite{ref:You05} and updated to show the domains allowed by both atomic physics measurements and PVES ones in 2007, either with or without the modification resulting from \cite{ref:Der09}.  The present agreement with the SM model is made conspicuous.
 There are two consequences. 1) Since the existence of an additional neutral gauge boson would alter the value of $Q_W^{ex}$ in a way depending on its mass, one can place a new limit on such an eventual $Z^{\prime}$ boson $M>1.3$ TeV which (in the frame of SO10 unification) turns out to be even higher than the limit obtained from a direct search persued at the Tevatron collider (0.82 TeV). 2) Using the new $Q_W^{ex}$ result one arrives at a determination of $sin^2{\theta^{eff}}= 0.2381(11)$, now agreeing with the SM, which is slightly more precise than the previously most precise low-energy test performed in the Moeller e-scattering experiment at SLAC, fig.\ref{PVES}.
 
 Present knowledge about the nuclear-spin contribution is quite limited: there is one sole result deduced from the same  Boulder experiment, in this case by comparing to 1 the ratio $r_{hf}$ of the $E_1^{pv}$ amplitudes measured on the two hyperfine $\Delta F= \pm 1$ hyperfine lines.  It includes three effects: the "anapole moment", the nucleon-axial contribution to the electroweak e-nucleon interaction and the perturbation of the nuclear-spin independent PV effect by the hyperfine interaction. The result obtained, $r_{hf}-1 = (5 \pm 0.7)\times 10^{-2}$ is actually very puzzling, since  
 the theoretical prediction \cite{ref:ana} is $ (1.6 \pm 0.3)\times 10^{-2}$. In addition, no such effect was observed in Tl in spite of the $1\%$   level of accuracy achieved.
 
 Finally, the recent observation of APV in Yb \cite{ref:Yb} has opened the route to a new experimental  approach.   This atom has 7 stable isotopes and it should be possible to measure the ratios of $E_1^{pv}$ for several of them. This makes  it possible to eliminate the atomic factor in the ratio and thus have access to  $Q_W$ isotopic effects for the first time, see D. Budker, this conference.   
 
 Today, there still remain very important goals to be achieved.

1) Measure $Q_W$ to 0.1$\%$ precision in Cesium in view of the obtained gain of precision
in atomic structure calculations. The Boulder result has to be cross-checked. 
Improvements are possible. The Paris 2005 exp (2.6$\%$ accurate) \cite{ref:exp2} was stopped 
by the powers that be, while its accuracy was still improving. 
Its principle is to take advantage of PV asymmetry amplification when a probe beam propagates through the vapor. Based on the experience acquired,  we have  suggested a different $E$-field configuration giving rise to a much larger  PV asymmetry amplification and  higher accuracy \cite{ref:pro}.
Another approach is to devise feasible experiments in francium and go ahead on the Ra$^{+}$ experiment. In both cases the effect is 20 times larger but atoms are scarce and one may expect atomic physics calculations will be achievable with an accuracy comparable to that obtained today in Cs (see S. Aubin and L. Willmann, this conference).

2) Design an experiment specifically sensitive to the nuclear spin-dependent PV effect, {\it i.e.} where its contribution  dominates that of $Q_W$: one has to solve the present puzzle raised by the Cs result. (For proposals, see sec. 3 and S. Cahn this conference.)

3) Make precise measurements of  $E_1^{pv}$ ratios for different Yb isotopes, to observe isotopic effects on $Q_W$  and possibly detect  the anapole moments  in  the odd isotopes.
\section{Prospects for APV measurements using matter-wave interferometry}
\begin{figure}
\centering\includegraphics[height=2.6cm]{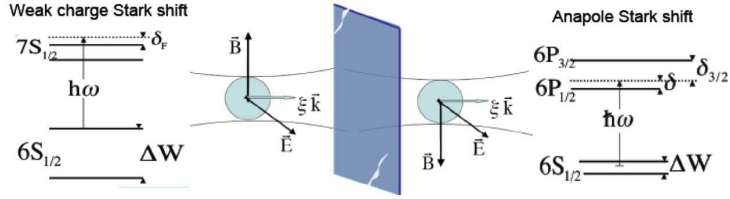}     
% includes {FigureBouchiat1}
\caption{Two dressed atoms of opposite handedness lead to opposite linear Stark shifts.}\label{chiral}
\end{figure}
 Let me now suggest new strategies for making APV measurements, inspired by the huge progress made in time and frequency metrology, over the past ten years. The field has been boosted by exploiting the methods of atom-interferometry, well-known for unprecedented accuracy. 
However, a conceptual difficulty arises: metrologists are used to measuring energy differences or frequency shifts, while all APV experiments so far have been based on L-R asymmetries in the transition rates  
 because there is no frequency shift associated with the PV  transition dipole. An electric dipole P-odd and T-even cannot give rise to a frequency shift in a stationnary atomic state perturbed by homogeneous E and B dc fields~\cite{ref:san}. This difficulty can be solved by using light shifts~\cite{ref:shift}.
\begin{figure}
\centering\includegraphics[height=3.8cm]{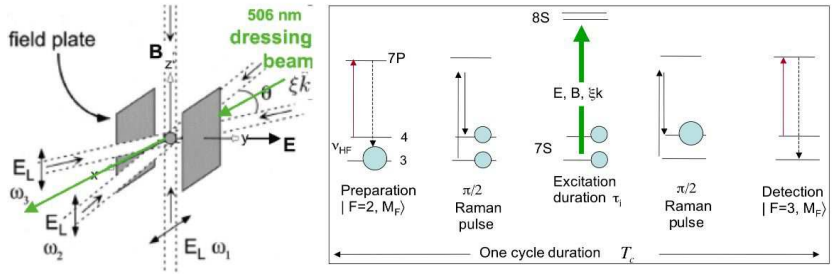}     
\caption{Left: Scheme for a PV Stark shift measurement inspired by a planned EDM project ({\it Adapted from\cite{ref:chin}}); Right: Sequence of measurement for Fr atoms. ({\it Courtesy of M.D. Plimmer})}. \label{EDM}
\end{figure}
  
When alkali atoms submitted to static electric and magnetic fields ($\vec E$, $\vec B$) are placed in a radiation field quasi-resonant for an atomic transition, a static  electric dipole moment (EDM) can appear, hence also a linear Stark shift arising from $V^{pv}$. For a circularly-polarized radiation field the signature of the shift is given by the pseudoscalar T-even quantity  $\chi = \vec E \wedge \xi \hat k \cdot B$, where $\xi$ denotes the field helicity and $\xi \hat k$ the photon angular momentum.  The mixed product specifies the handedness of the field configuration. There are actually two kinds of linear Stark shifts, depending on the light frequency. If the dressing beam is detuned with respect to the highly forbidden transition, the linear Stark shift involves the nuclear weak charge. In the case of a detuning with respect to the resonance transition ({\it e.g.} $6S_{1/2}-6P_{1/2}$ in Cs), the shift is largely dominated by the nuclear-spin dependent PV effect.
In both cases we have shown that dressed alkali atoms can thus be artificially endowed with handedness \cite{ref:shift}. Two atoms of opposite handedness  (fig.\ref{chiral}) behave similarly to enantiomer molecules \cite{ref:Bor}. 
The handedness of the field configuration plays the same  role as that of the chemical site inside an enantiomer molecule. For two mirror-image atoms one expects opposite shifts, just like for mirror-image molecules.
However, there is a price to be paid: when the ground-state is admixed with the excited state by the dressing beam, the cold atom cloud acquires a certain decay time. One must ensure that over the time interval needed for observing the linear Stark shift of the hyperfine or Zeeman transition frequencies, the dressed ground state decay can be neglected.
Therefore, the dressing beam intensity, its detuning and the electric field magnitude have to fulfil compromises discussed in~\cite{ref:shift}. 

We illustrate our proposal on a design planned to search for a static T-odd EDM shift, with state of the art methods of atomic interferometry \cite{ref:chin}, fig.\ref{EDM}. We assume an interaction time $\tau_i$= 1 s and a cycling time of 2s. For $^{221}$Fr, $I=5/2$, we predict  a weak-charge Stark shift of 100 $\mu$Hz for a dressing beam intensity of 10 kW/cm$^2$, $E \approx $200 V/cm, detuning $\delta_F/2\pi \approx$ 130 MHz. Measurements have to be performed in combination  with parameter reversal and calibration sequences.  The light shift and intensity noise from the dressing beam can be eliminated. 
For a sample of $N_{at}=10^{6}$ cold Fr atoms, in conditions of projection-noise limited sensitivity, the signal to noise ($\propto \sqrt {N_{at}\,t}$) reaches 30 after a one hour averaging time, $t$.   
In the case of the anapole Stark shift, since the dressing beam is detuned from an allowed transition, the stability condition no longer involves the magnitude of $E$, which can therefore be made as large as possible. With the design of fig.\ref{EDM}, the same dressing beam intensity, $E$=100 kV/cm and optimized detuning (fig.\ref{chiral}),
one expects an anapole Stark shift $\sim 400 \; \mu$Hz for Fr ($\sim$10 times less for Cs).
\acknowledgments
L. Pottier, J. Gu\'ena, Ph. Jacquier, M. Lintz, and other team members who cannot all be listed here, have made  invaluable contributions to the field of  atomic parity violation.  I express my warmest thanks to all of them.

\end{document}